\documentclass[journal]{IEEEtran}

\usepackage{graphicx,latexsym,color}
\usepackage{amsmath,amsthm,amsfonts}
\usepackage{latexsym}
\usepackage{cite,citesort}
\usepackage[normalem]{ulem}	

\definecolor{MyDarkBlue}{rgb}{0.1,0,0.55} 
\definecolor{MyRed}{rgb}{1,0,0} 
\definecolor{MyBlue}{rgb}{0,0,1} 
\definecolor{MyGreen}{rgb}{0,1,0} 

\usepackage{bm}
\usepackage{tikz}
\usepackage{pgfplots}
\usetikzlibrary{mindmap,trees}
\usetikzlibrary{%
  3d,
  calc,
  arrows,%
  shapes.misc,
  shapes.arrows,%
  chains,%
  matrix,%
  positioning,
  scopes,%
  decorations.pathmorphing,
  shadows%
}

\renewcommand{\vec}[1]{{\boldsymbol#1}}
\newcommand{\R}{\mathbb{R}}

\newcommand{\Z}{\mathbb{Z}}

\newcommand{\ie}{\textit{i.e.}\/, }
\newcommand{\eg}{\textit{e.g.}\/, }
\newcommand{\cf}{\textit{cf.}\/, }

\providecommand*{\mrm}[1]{\mathrm{#1}}
\providecommand*{\unit}[1]{\ensuremath{\mrm{\,#1}}}
\providecommand*{\eu}{\ensuremath{\mrm{e}}}
\providecommand*{\iu}{\ensuremath{\mrm{i}}}

\providecommand*{\diff}{\operatorname{d}\!}
\renewcommand{\Im}{\operatorname{Im}}	

\begin{document}

\title{On the natural modes of helical structures}

\author{Sven~Nordebo,~\IEEEmembership{Senior Member,~IEEE,}
        Mats~Gustafsson,~\IEEEmembership{Member,~IEEE,}
        Gerhard~Kristensson,~\IEEEmembership{Senior Member,~IEEE,}
        B\"{o}rje~Nilsson,
        Alexander~Nosich,~\IEEEmembership{Fellow,~IEEE,}
        and~Daniel~Sj\"{o}berg,~\IEEEmembership{Member,~IEEE}
        \thanks{Manuscript received \today. This work was supported in part by the Swedish Foundation for Strategic Research (SSF).}  
\thanks{S. Nordebo is with the Department of Physics and Electrical Engineering, Linn\ae us University, 
351 95 V\"{a}xj\"{o}, Sweden. Phone: +46 470 70 8193. Fax: +46 470 84004. E-mail: sven.nordebo@lnu.se.}
\thanks{M. Gustafsson is with the Department of Electrical and Information Technology, Lund University, Box 118, 
221 00 Lund, Sweden. Phone: +46 46 222 7506. Fax:  +46 46 129948. E-mail:  mats.gustafsson@eit.lth.se.}
\thanks{G. Kristensson is with the Department of Electrical and Information Technology, Lund University, Box 118, 
221 00 Lund, Sweden. Phone: +46 46 2224562. Fax:  +46 46 2227508. E-mail:  gerhard.kristensson@eit.lth.se.},
\thanks{B. Nilsson is with the Department of Mathematics, Linn\ae us University, 
351 95 V\"{a}xj\"{o}, Sweden. Phone: +46 470 70 8849. Fax:  +46 470 84004. E-mail:  borje.nilsson@lnu.se.}
\thanks{A. Nosich is with the Institute of Radio Physics and Electronics of the National Academy of Sciences of Ukraine (IRE NASU), 
vul. Proskury 12, Kharkiv 61085, Ukraine. E-mail: anosich@yahoo.com.}
\thanks{D. Sj\"{o}berg is with the Department of Electrical and Information Technology, Lund University, Box 118, 
221 00 Lund, Sweden. Phone: +46 46 222 7511. Fax:  +46 46 129948. E-mail:  daniel.sjoberg@eit.lth.se.}
}



\maketitle

\begin{abstract}
Natural modes of helical structures are treated by using the periodic dyadic Green's functions in cylindrical coordinates.
The formulation leads to an infinite system of one-dimensional integral equations in reciprocal (Fourier) space. 
Due to the twisted structure of the waveguide together with a quasi-static assumption
the set of non-zero coefficients in reciprocal space is sparse and the formulation can therefore be used 
in a numerical method based on a truncation of the set of coupled integral equations.
The periodic dyadic Green's functions are furthermore useful in a simple direct calculation of
the quasi-static fields generated by 
thin helical wires.
\end{abstract}


\section{Introduction}\label{sect:introduction}
\IEEEPARstart{T}{}he purpose of this paper is to formulate a volume integral equation for the determination of
the natural modes of helical structures. Low-frequency applications are of particular importance 
where it can be anticipated the existence of twisted modes of a particularly simple structure.
The presented problem formulation is largely motivated by the need of being able to accurately
model the field distribution and losses inside twisting three-phase high-voltage power cables at 50\unit{Hz}, see \eg \cite{Gustavsen+etal2009,Habib+Kordi2013}.
The approach could also potentially be useful for analyzing the wave propagation characteristics of the so called litz wires.

Helical waveguide structures have been treated previously such as \eg with helical sheaths \cite{Collin1991,Morgan+Young1956},
and approximations for wire helices\cite{Sensiper1955,Moser+Spencer1968}. Presently, there are also very promising numerical techniques being developed
that are based on Finite Element Modeling \cite{Gustavsen+etal2009,Habib+Kordi2013} and the Method of Moments \cite{Patel+etal2013}.
However, to our knowledge
there has not been any general presentation regarding analytical modeling of the natural modes of helical structures.
It is the aim of this paper to fill in this gap.
On the other hand, there is a large body of literature on the general dispersion properties of open waveguides, see \eg
\cite{Nosich1994,Nosich2000,Kartchevski2000,Kartchevski+etal2005,Rozzi+Mongiardo1997},
as well as on the general properties of the electromagnetic volume integral equations, see \eg 
\cite{Botha2006,Budko+Samokhin2006,Chew1995,Costabel+etal2010,Sancer+etal2006,Volakis+Sertel2011}.
In particular, it is well known that the volume integral operators in electromagnetics are strongly singular
and that many questions regarding their spectral theory remain largely open \cite{Costabel+etal2010}. 
Nevertheless, it can be shown in very general settings that modes of open waveguides exist and can 
be interpreted in terms of poles of a meromorphic Fourier transform and that these poles depend continuously
on the model data (except for points where poles coalesce or at the boundary of the domain of meromorphicity, 
\ie at infinity and at the branch-point corresponding to the wavenumber of the exterior domain) \cite{Nosich2000}. 
When there are sources present, the modes (the discrete set of eigenfunctions) can be obtained as the residues 
of the poles and the non-discrete set is manifested as an integration along the branch-cut \cite{Collin1991}. 
In practical circumstances, the branch-cut contribution can often be neglected \cite{Nordebo+etal2014a}.

In this paper, a general helical waveguide structure is treated by using classical analytic function theory and 
Fourier techniques (in particular the convolution theorem) in connection with the electromagnetic volume integral formulation 
and a cylindrical vector wave expansion of the related dyadic Green's functions.
The Floquet modes are defined as the poles of the corresponding integral operators,
and the analytic periodic Green's functions are derived by employing the classical Poisson summation formula.
Two independent approaches (with and without explicit sources) are used to derive the resulting 
infinite system of one-dimensional integral equations. These equations can then be discretized by truncation and 
by using a standard collocation method.

As a useful byproduct, the periodic dyadic Green's functions provide a simple direct calculation of
the quasi-static fields generated by thin helical wires. 
\section{Natural modes with finite sources}

\subsection{Preliminaries}

Consider a straight helical waveguide structure of radius $a$ consisting of a twisting
inhomogeneous and anisotropic 
material. 
The twisting means that the cross-section of the guide is rotating along the longitudinal direction of the structure.
The waveguide may consist of several layers with different twist, but it is assumed that the material has
a smallest common period $p$ in the longitudinal direction.
The waveguide is assumed to be lossy, and it constitutes an open structure placed in a surrounding homogeneous and isotropic 
free space.

Let $\mu_0$, $\epsilon_0$, $\eta_0$ and ${\rm c}_0$ denote the permeability, the permittivity, the wave impedance and
the speed of light in vacuum, respectively, and where $\eta_0=\sqrt{\mu_0/\epsilon_0}$ and ${\rm c}_0=1/\sqrt{\mu_0\epsilon_0}$.
The wavenumber of vacuum is given by $k_0=\omega\sqrt{\mu_0\epsilon_0}$ where $\omega=2\pi f$ is the angular frequency and $f$ the frequency.
The relative permeability and permittivity of the surrounding free space are denoted by $\mu$ and $\epsilon$, respectively, and the corresponding
wavenumber is given by $k=k_0\sqrt{\mu\epsilon}$.
The cylindrical coordinates are denoted by $(\rho,\phi,z)$, the corresponding unit vectors $(\hat{\vec{\rho}},\hat{\vec{\phi}},\hat{\vec{z}})$,
the transversal coordinate vector $\vec{\rho}=\rho\hat{\vec{\rho}}$ and the radius vector $\vec{r}=\vec{\rho}+z\hat{\vec{z}}$.  

Let $\vec{E}(\vec{r})$ and $\vec{H}(\vec{r})$ denote the electric and magnetic fields, respectively, 
where the time-harmonic factor  $\eu^{-\iu\omega t}$ has been suppressed. Further, let $\vec{J}_{\rm s}(\vec{r})$ and $\vec{M}_{\rm s}(\vec{r})$ denote
the imposed electric and magnetic sources, respectively, and which are assumed to be constrained to a finite region $V_{\rm s}$ inside the waveguide structure.
Maxwell's equations \cite{Jackson1999} for this situation are given by
\begin{equation}\label{eq:Maxwell1}
\left\{\begin{array}{l}
\nabla\times \vec{E}(\vec{r})=\iu\omega\mu_0\vec{\mu}(\vec{r})\cdot\vec{H}(\vec{r})-\vec{M}_{\rm s}(\vec{r}), \vspace{0.2cm}\\
\nabla\times\vec{H}(\vec{r})=-\iu\omega\epsilon_0\vec{\epsilon}(\vec{r})\cdot\vec{E}(\vec{r})+\vec{J}_{\rm s}(\vec{r}),
\end{array}\right.
\end{equation}
where $\vec{\mu}(\vec{r})$ and $\vec{\epsilon}(\vec{r})$ are the complex valued relative permeability and permittivity dyadics of the material, respectively.
The Maxwell's equations \eqref{eq:Maxwell1} are also supplemented with
a radiation condition providing a unique solution with fields vanishing at infinity \cite{Chew1995}. 
The equations \eqref{eq:Maxwell1} can be reformulated in terms of the surrounding free space as
\begin{equation}\label{eq:Maxwell1eqsource}
\left\{\begin{array}{l}
\nabla\times \vec{E}(\vec{r})=\iu\omega\mu_0\mu\vec{H}(\vec{r})-\vec{M}(\vec{r})-\vec{M}_{\rm s}(\vec{r}), \vspace{0.2cm}\\
\nabla\times\vec{H}(\vec{r})=-\iu\omega\epsilon_0\epsilon\vec{E}(\vec{r})+\vec{J}(\vec{r})+\vec{J}_{\rm s}(\vec{r}),
\end{array}\right.
\end{equation}
where $\vec{M}(\vec{r})$ and $\vec{J}(\vec{r})$ are the equivalent  magnetic and electric sources 
\begin{equation}\label{eq:eqsourcedef}
\left\{\begin{array}{l}
\vec{M}(\vec{r})=-\iu\omega\mu_0\mu\vec{\chi}_{\rm m}(\vec{r})\cdot\vec{H}(\vec{r}), \vspace{0.2cm}\\
\vec{J}(\vec{r})=-\iu\omega\epsilon_0\epsilon\vec{\chi}_{\rm e}(\vec{r})\cdot\vec{E}(\vec{r}),
\end{array}\right.
\end{equation}
where $\vec{\chi}_{\rm m}(\vec{r})$ and $\vec{\chi}_{\rm e}(\vec{r})$ are the magnetic and electric susceptibility (or contrast) dyadics, respectively,
defined by
\begin{equation}\label{eq:ximxiedef}
\left\{\begin{array}{l}
\vec{\chi}_{\rm m}(\vec{r})=\frac{1}{\mu}\vec{\mu}(\vec{r})-\vec{I}, \vspace{0.2cm}\\
\vec{\chi}_{\rm e}(\vec{r})=\frac{1}{\epsilon}\vec{\epsilon}(\vec{r})-\vec{I},
\end{array}\right.
\end{equation}
and where $\vec{I}$ is the identity dyadic.

The solution to \eqref{eq:Maxwell1} can now be expressed in terms of the following integral equation of the second kind
\begin{equation}\label{eq:inteq1}
\left\{\begin{array}{l}
\vec{E}(\vec{r})-k^2\int_{V_{\rm m}}\vec{G}_{\rm e}(\vec{r},\vec{r}^\prime,k)\cdot\vec{\chi}_{\rm e}(\vec{r}^\prime)\cdot\vec{E}(\vec{r}^\prime)\diff v^\prime \vspace{0.2cm}\\
\hspace{1cm} -\iu\omega\mu_0\mu\int_{V_{\rm m}}\vec{G}_{\rm m}(\vec{r},\vec{r}^\prime,k)\cdot\vec{\chi}_{\rm m}(\vec{r}^\prime)\cdot\vec{H}(\vec{r}^\prime)\diff v^\prime \vspace{0.2cm}\\
\hspace{2cm}=\iu\omega\mu_0\mu\int_{V_{\rm s}}\vec{G}_{\rm e}(\vec{r},\vec{r}^\prime,k)\cdot\vec{J}_{\rm s}(\vec{r}^\prime)\diff v^\prime \vspace{0.2cm}\\
\hspace{3cm}-\int_{V_{\rm s}}\vec{G}_{\rm m}(\vec{r},\vec{r}^\prime,k)\cdot\vec{M}_{\rm s}(\vec{r}^\prime)\diff v^\prime, \vspace{0.2cm}\\
\vec{H}(\vec{r})+\iu\omega\epsilon_0\epsilon\int_{V_{\rm m}}\vec{G}_{\rm m}(\vec{r},\vec{r}^\prime,k)\cdot\vec{\chi}_{\rm e}(\vec{r}^\prime)\cdot\vec{E}(\vec{r}^\prime)\diff v^\prime \vspace{0.2cm}\\
\hspace{1cm} -k^2\int_{V_{\rm m}}\vec{G}_{\rm e}(\vec{r},\vec{r}^\prime,k)\cdot\vec{\chi}_{\rm m}(\vec{r}^\prime)\cdot\vec{H}(\vec{r}^\prime)\diff v^\prime \vspace{0.2cm}\\
\hspace{2cm}=\int_{V_{\rm s}}\vec{G}_{\rm m}(\vec{r},\vec{r}^\prime,k)\cdot\vec{J}_{\rm s}(\vec{r}^\prime)\diff v^\prime \vspace{0.2cm}\\
\hspace{3cm} +\iu\omega\epsilon_0\epsilon\int_{V_{\rm s}}\vec{G}_{\rm e}(\vec{r},\vec{r}^\prime,k)\cdot\vec{M}_{\rm s}(\vec{r}^\prime)\diff v^\prime, 
\end{array}\right.
\end{equation}
where the integrals extend over the support of the material $V_{\rm m}=\{\vec{r}|(\rho,\phi,z)\in[0,a]\times[0,2\pi]\times[-\infty,+\infty]\}$, and the finite source region $V_{\rm s}$, respectively,
and where $\vec{r}\in V_{\rm m}$, \cf \cite{Botha2006,Budko+Samokhin2006,Collin1991,Chew1995,Costabel+etal2010,Sancer+etal2006}.
Here, $\vec{G}_{\rm e}(\vec{r},\vec{r}^\prime,k)$ and $\vec{G}_{\rm m}(\vec{r},\vec{r}^\prime,k)$ are the electric and magnetic dyadic 
Green's functions for the surrounding free space, respectively, defined by
\begin{equation}\label{eq:EMdyaddef}
\left\{\begin{array}{l}
\vec{G}_{\rm e}(\vec{r},\vec{r}^\prime,k)=\{\vec{I}+\frac{1}{k^2}\nabla\nabla\} G(\vec{r},\vec{r}^\prime,k), \vspace{0.2cm}\\
\vec{G}_{\rm m}(\vec{r},\vec{r}^\prime,k)=\nabla G(\vec{r},\vec{r}^\prime,k)\times\vec{I},
\end{array}\right.
\end{equation}
and where $G(\vec{r},\vec{r}^\prime,k)=\frac{\eu^{\iu k|\vec{r}-\vec{r}^\prime|}}{4\pi|\vec{r}-\vec{r}^\prime|}$
is the corresponding scalar Green's function,  see \eg \cite{Collin1991,Chew1995,Jin2010}.

In cylindrical coordinates, the electric dyadic Green's function $\vec{G}_{\rm e}(\vec{r},\vec{r}^\prime,k)$ can be expressed as
\begin{equation}\label{eq:Ge0exp0}
\vec{G}_{\rm e}(\vec{r},\vec{r}^\prime,k)=\vec{G}_{\rm e}^0(\vec{r},\vec{r}^\prime,k)-\frac{1}{k^2}\hat{\vec{\rho}}\hat{\vec{\rho}}\delta(\vec{r}-\vec{r}^\prime),
\end{equation}
where $\vec{G}_{\rm e}^0(\vec{r},\vec{r}^\prime,k)$ is the part of $\vec{G}_{\rm e}(\vec{r},\vec{r}^\prime,k)$ 
which can be expanded in transverse (solenoidal) cylindrical vector waves for $\vec{r}\neq\vec{r}^\prime$,
$\delta(\vec{r}-\vec{r}^\prime)$ the three-dimensional delta distribution and
$-\frac{1}{k^2}\hat{\vec{\rho}}\hat{\vec{\rho}}$ the corresponding source-point dyadics \cite{Collin1991,Chew1995}. 
Here, the expansion in cylindrical vector waves is given by
\begin{multline}\label{eq:Ge0exp}
\vec{G}_{\rm e}^0(\vec{r},\vec{r}^\prime,k)\\
=\frac{1}{2\pi}\int_{-\infty}^{\infty}\sum_{m=-\infty}^{\infty}
\vec{a}_m(\vec{\rho},\vec{\rho}^\prime,k,\alpha)\eu^{\iu m(\phi-\phi^\prime)}\eu^{\iu \alpha(z-z^\prime)}\diff\alpha,
\end{multline}
where the dyadic $\vec{a}_m(\vec{\rho},\vec{\rho}^\prime,k,\alpha)$ is defined in \eqref{eq:dyada} in the appendix~\ref{sect:modeexp}. 
Similarly, the magnetic dyadic Green's function $\vec{G}_{\rm m}(\vec{r},\vec{r}^\prime,k)$ is expanded in cylindrical vector waves based on the dyadic $\vec{b}_m(\vec{\rho},\vec{\rho}^\prime,k,\alpha)$ 
defined in \eqref{eq:dyadb}. Note that $\vec{G}_{\rm m}(\vec{r},\vec{r}^\prime,k)$ 
does not contain any source-point dyadics at $\vec{r}=\vec{r}^\prime$ \cite{Yaghjian1980}.

To simplify the description below, a non-magnetic material is assumed where $\vec{\chi}_{\rm m}(\vec{r})=\vec{0}$ and $\vec{\chi}_{\rm e}(\vec{r})=\vec{\chi}(\vec{r})$, 
and only the first equation is needed in \eqref{eq:inteq1}.
A generalization to the full system \eqref{eq:inteq1} will be straightforward.
Extracting the contribution from the source-point as defined in \eqref{eq:Ge0exp0}, the electric field integral equation in \eqref{eq:inteq1} can now be expressed as
\begin{multline}\label{eq:inteq2}
\left[\vec{I}+\hat{\vec{\rho}}\hat{\vec{\rho}}\cdot\vec{\chi}(\vec{r})\right]\cdot\vec{E}(\vec{r}) \\
-k^2\int_{V_{\rm m}}\vec{G}_{\rm e}^0(\vec{r},\vec{r}^\prime,k)\cdot\vec{\chi}(\vec{r}^\prime)\cdot\vec{E}(\vec{r}^\prime)\diff v^\prime 
=\vec{F}(\vec{r}), 
\end{multline}
where $\vec{r}\in V_{\rm m}$, and the source vector function $\vec{F}(\vec{r})$ is given by
\begin{multline}\label{eq:Fdef}
\vec{F}(\vec{r})=\iu\omega\mu_0\mu\int_{V_{\rm s}}\vec{G}_{\rm e}^0(\vec{r},\vec{r}^\prime,k)\cdot\vec{J}_{\rm s}(\vec{r}^\prime)\diff v^\prime \\
-\iu\omega\mu_0\mu\frac{1}{k^2}\hat{\vec{\rho}}\hat{\vec{\rho}}\cdot\vec{J}_{\rm s}(\vec{r})
-\int_{V_{\rm s}}\vec{G}_{\rm m}(\vec{r},\vec{r}^\prime,k)\cdot\vec{M}_{\rm s}(\vec{r}^\prime)\diff v^\prime.
\end{multline}

At each radial distance $\rho$ the material is periodic in the $\phi$-coordinate with period $2\pi$ 
and in the $z$-coordinate with period $p$. The material dyadic $\vec{\chi}(\vec{r})$ can hence be represented by the following two-dimensional Fourier series
\begin{equation}\label{eq:chidef}
\vec{\chi}(\vec{r})=\displaystyle\sum_{m=-\infty}^{\infty}\sum_{n=-\infty}^{\infty}\vec{\chi}_{mn}(\vec{\rho})\eu^{\iu m\phi}\eu^{\iu n\frac{2\pi}{p}z}, 
\end{equation}
where $\vec{\chi}_{mn}(\vec{\rho})$ are the corresponding dyadic Fourier series coefficients.

\subsection{Fourier analysis}

Consider the following Fourier representation of the electric field 
\begin{equation}\label{eq:EdefF}
\vec{E}(\vec{r})=\frac{1}{2\pi}\int_{-\infty}^{\infty}\sum_{m=-\infty}^{\infty}\vec{E}_m(\vec{\rho},\alpha)\eu^{\iu m\phi}\eu^{\iu\alpha z}\diff\alpha,
\end{equation}
and similarly for the sources $\vec{J}_{\rm s}(\vec{r})$ and $\vec{M}_{\rm s}(\vec{r})$ and the
source vector function $\vec{F}(\vec{r})$. The material dyadic is furthermore represented as
\begin{equation}\label{eq:chidefF}
\vec{\chi}(\vec{r})=\frac{1}{2\pi}\int_{-\infty}^{\infty}\sum_{m=-\infty}^{\infty}\vec{\chi}_m(\vec{\rho},\alpha)\eu^{\iu m\phi}\eu^{\iu\alpha z}\diff\alpha,
\end{equation}
where
\begin{equation}\label{eq:chidefF2}
\vec{\chi}_m(\vec{\rho},\alpha)=2\pi\sum_{n=-\infty}^{\infty}\vec{\chi}_{mn}(\vec{\rho})\delta(\alpha-n\frac{2\pi}{p}),
\end{equation}
and where the coefficients $\vec{\chi}_{mn}(\vec{\rho})$ have been defined in \eqref{eq:chidef}.

Based on the convolution theorem the Fourier transformation of \eqref{eq:inteq2} yields the following integral equation in Fourier space
\begin{multline}\label{eq:inteqFspace1}
\vec{E}_m(\vec{\rho},\alpha)\\
+\hat{\vec{\rho}}\hat{\vec{\rho}}\cdot\frac{1}{2\pi}\int_{-\infty}^{\infty}\sum_{m^\prime=-\infty}^{\infty}
\vec{\chi}_{m-m^\prime}(\vec{\rho},\alpha-\alpha^\prime)\cdot \vec{E}_{m^\prime}(\vec{\rho},\alpha^\prime)\diff\alpha^\prime \\
- k^2 2\pi\int_{0}^a\vec{a}_m(\vec{\rho},\vec{\rho}^\prime,k,\alpha)
\cdot \frac{1}{2\pi}\int_{-\infty}^{\infty}\sum_{m^\prime=-\infty}^{\infty}\vec{\chi}_{m-m^\prime}(\vec{\rho}^\prime,\alpha-\alpha^\prime) \\ \cdot \vec{E}_{m^\prime}(\vec{\rho}^\prime,\alpha^\prime)\diff\alpha^\prime
\rho^\prime\diff\rho^\prime 
=\vec{F}_{m}(\vec{\rho},\alpha),
\end{multline}
where $\rho\in[0,a]$, $m\in\Z$ and $\alpha\in\R$.
The Fourier transformation of the source vector function is given by
\begin{multline}\label{eq:Fmrhoalpha}
\vec{F}_{m}(\vec{\rho},\alpha)=
\iu\omega\mu_0\mu 2\pi\int_{0}^{a}\vec{a}_m(\vec{\rho},\vec{\rho}^\prime,k,\alpha)\cdot\vec{J}_{{\rm s}m}(\vec{\rho}^\prime,\alpha)\rho^\prime\diff\rho^\prime\\
-\iu\omega\mu_0\mu\frac{1}{k^2}\hat{\vec{\rho}}\hat{\vec{\rho}}\cdot\vec{J}_{{\rm s}m}(\vec{\rho},\alpha) \\
-2\pi\int_{0}^{a}\vec{b}_m(\vec{\rho},\vec{\rho}^\prime,k,\alpha)\cdot\vec{M}_{{\rm s}m}(\vec{\rho}^\prime,\alpha)\rho^\prime\diff\rho^\prime.
\end{multline}
By further exploiting the distributional property given in \eqref{eq:chidefF2}, the integral equation \eqref{eq:inteqFspace1} becomes
\begin{multline}\label{eq:inteqFspace2}
\vec{E}_m(\vec{\rho},\alpha)+\hat{\vec{\rho}}\hat{\vec{\rho}}
\cdot\sum_{m^\prime=-\infty}^{\infty}\sum_{l=-\infty}^{\infty}
\vec{\chi}_{m-m^\prime,l}(\vec{\rho})\cdot \vec{E}_{m^\prime}(\vec{\rho},\alpha-l\frac{2\pi}{p}) \\
- k^2 2\pi\int_{0}^a\vec{a}_m(\vec{\rho},\vec{\rho}^\prime,k,\alpha)
\cdot\sum_{m^\prime=-\infty}^{\infty}\sum_{l=-\infty}^{\infty}\vec{\chi}_{m-m^\prime,l}(\vec{\rho}^\prime) \\ \cdot \vec{E}_{m^\prime}(\vec{\rho}^\prime,\alpha-l\frac{2\pi}{p})
\rho^\prime\diff\rho^\prime =\vec{F}_{m}(\vec{\rho},\alpha).
\end{multline}

The integral operator defined on the left-hand side of \eqref{eq:inteqFspace2} can be continued to the complex 
$\alpha$-plane taking the branch-cut (defined by the square root $\kappa=\sqrt{k^2-\alpha^2}$)
into proper account, and hence defines an analytic operator-valued function $A(\alpha)$ where $A(\alpha)\vec{E}(\vec{\rho},\alpha)=\vec{F}(\vec{\rho},\alpha)$.
The solution to \eqref{eq:inteqFspace2} can be expressed as the sum of discrete modes (residues at poles), 
plus an integration along the branch-cut \cite{Collin1991}. 
In practice, the branch-cut contribution can often be neglected \cite{Nordebo+etal2014a}.



\subsection{The Floquet theorem}

It is assumed that the sources $\vec{J}_{\rm s}(\vec{r})$ and $\vec{M}_{\rm s}(\vec{r})$ are supported only for $z<0$, and hence that the 
source vector function $\vec{F}_{m}(\vec{\rho},\alpha)$ defined in \eqref{eq:Fmrhoalpha} is an analytic function in the upper half-plane $\Im\alpha>0$.
According to the Floquet theorem \cite{Collin1991},  there will be Floquet modes 
in the source-free region $z>0$, and where the electric field can be expressed as
\begin{equation}\label{eq:Floquetmodedef}
\vec{E}(\vec{r})=\sum_{m=-\infty}^{\infty}\sum_{n=-\infty}^{\infty}\vec{E}_{mn}(\vec{\rho})\eu^{\iu m\phi}\eu^{\iu n\frac{2\pi}{p}z}\eu^{\iu\beta z},
\end{equation}
where $\beta$ is the complex valued propagation constant  of the Floquet mode and where $\Im\beta>0$.
The Fourier transform $\vec{E}_m(\vec{\rho},\alpha)$ can hence be written as
\begin{equation}
\vec{E}_m(\vec{\rho},\alpha)=\sum_{n=-\infty}^{\infty}\vec{E}_{mn}(\vec{\rho})\frac{1}{\iu(\alpha-(\beta+n\frac{2\pi}{p}))}+\vec{E}_m^{-}(\vec{\rho},\alpha),
\end{equation}
with poles at $\beta+n\frac{2\pi}{p}$. Here, $\vec{E}_m^{-}(\vec{\rho},\alpha)$ is the Fourier transform of the left-sided part of
$\vec{E}(\vec{r})$ which is supported in $z<0$, and hence $\vec{E}_m^{-}(\vec{\rho},\alpha)$ is analytic in the upper half-plane $\Im\alpha>0$.

The residue theorem can be stated here as
\begin{equation}\label{eq:residuetheorem}
\frac{1}{2\pi}\oint_{{\cal C}_n}\frac{f(\alpha)\diff\alpha}{\iu(\alpha-(\beta+q\frac{2\pi}{p}))}=f(\beta+n\frac{2\pi}{p})\delta_{nq},
\end{equation}
where ${\cal C}_n$ is a small closed contour enclosing the point $\beta+n\frac{2\pi}{p}$,
$f(\alpha)$ is a function which is analytic inside ${\cal C}_n$ and $\delta_{nq}$ denotes the Kronecker delta.
By applying the residue theorem \eqref{eq:residuetheorem} to \eqref{eq:inteqFspace2}, the following integral
equation is obtained for $\vec{E}_{mn}(\vec{\rho})$
\begin{multline}\label{eq:integraleqFinal}
\vec{E}_{mn}(\vec{\rho})+\hat{\vec{\rho}}\hat{\vec{\rho}}\cdot\sum_{m^\prime=-\infty}^{\infty}\sum_{n^\prime=-\infty}^{\infty}
\vec{\chi}_{m-m^\prime,n-n^\prime}(\vec{\rho})\cdot\vec{E}_{m^\prime n^\prime}(\vec{\rho})\\
-k^2 2\pi \int_{0}^a\vec{a}_m(\vec{\rho},\vec{\rho}^\prime,k,\beta+n\frac{2\pi}{p})\\
\cdot\sum_{m^\prime=-\infty}^{\infty}\sum_{n^\prime=-\infty}^{\infty}\vec{\chi}_{m-m^\prime,n-n^\prime}(\vec{\rho}^\prime)
\cdot\vec{E}_{m^\prime n^\prime}(\vec{\rho}^\prime)\rho^\prime\diff\rho^\prime=\vec{0},
\end{multline}
where $\rho\in[0,a]$ and $(m,n)$ are integers.

\section{Natural modes with periodic excitation}

\subsection{Preliminaries}
Maxwell's equations for the free space with a periodically modulated plane wave excitation can be written
\begin{equation}\label{eq:Maxwell2per}
\left\{\begin{array}{l}
\nabla\times \vec{E}(\vec{r})=\iu\omega\mu_0\mu\vec{H}(\vec{r})-\vec{M}(\vec{r}), \vspace{0.2cm}\\
\nabla\times\vec{H}(\vec{r})=-\iu\omega\epsilon_0\epsilon\vec{E}(\vec{r})+\vec{J}(\vec{r}),
\end{array}\right.
\end{equation}
where the magnetic source is given by $\vec{M}(\vec{r})=\widetilde{\vec{M}}(\vec{r})\eu^{\iu\beta z}$
with a $p$-periodic part $\widetilde{\vec{M}}(\vec{r})=\widetilde{\vec{M}}(\vec{r}+\hat{\vec{z}}p)$, and similarly
for the electric source $\vec{J}(\vec{r})=\widetilde{\vec{J}}(\vec{r})\eu^{\iu\beta z}$, and where $\beta$ is the Floquet wavenumber 
with $\Im\beta\geq 0$. In the transverse plane the sources are assumed to be constrained to the cross-sectional area $S$ with radius $a$. 
The Maxwell's equations \eqref{eq:Maxwell2per} are also supplemented with a radiation condition in the transverse plane.

Following the standard derivation based on vector potentials as in \eg \cite{Jin2010}, the solution to \eqref{eq:Maxwell2per}
can be written
\begin{equation}\label{eq:Maxwell2persol}
\left\{\begin{array}{l}
\vec{E}(\vec{r})=\iu\omega\mu_0\mu\int_{S}\int_{0}^p\vec{G}_{\rm ep}(\vec{r},\vec{r}^\prime,k,\beta)\cdot\vec{J}(\vec{r}^\prime)\diff S^\prime\diff z^\prime \vspace{0.2cm}\\
\hspace{2cm}-\int_{S}\int_{0}^p\vec{G}_{\rm mp}(\vec{r},\vec{r}^\prime,k,\beta)\cdot\vec{M}(\vec{r}^\prime)\diff S^\prime\diff z^\prime, \vspace{0.2cm}\\
\vec{H}(\vec{r})=\int_{S}\int_{0}^p\vec{G}_{\rm mp}(\vec{r},\vec{r}^\prime,k,\beta)\cdot\vec{J}(\vec{r}^\prime)\diff S^\prime\diff z^\prime \vspace{0.2cm}\\
\hspace{1.5cm} +\iu\omega\epsilon_0\epsilon\int_{S}\int_{0}^p\vec{G}_{\rm ep}(\vec{r},\vec{r}^\prime,k,\beta)\cdot\vec{M}(\vec{r}^\prime)\diff S^\prime\diff z^\prime, 
\end{array}\right.
\end{equation}
where the integration is over one unit cell $V_{\rm c}=\{\vec{r}|(\vec{\rho},z)\in S\times [0,p]$\} and $\vec{r}\in V_{\rm c}$, 
and where $\vec{G}_{\rm ep}(\vec{r},\vec{r}^\prime,k,\beta)$ and $\vec{G}_{\rm mp}(\vec{r},\vec{r}^\prime,k,\beta)$
are the periodic electric and magnetic Green's dyadics, respectively, defined by
\begin{equation}\label{eq:perEMdyaddef}
\left\{\begin{array}{l}
\vec{G}_{\rm ep}(\vec{r},\vec{r}^\prime,k,\beta)=\{\vec{I}+\frac{1}{k^2}\nabla\nabla\} G_{\rm p}(\vec{r},\vec{r}^\prime,k,\beta), \vspace{0.2cm}\\
\vec{G}_{\rm mp}(\vec{r},\vec{r}^\prime,k,\beta)=\nabla G_{\rm p}(\vec{r},\vec{r}^\prime,k,\beta)\times\vec{I},
\end{array}\right.
\end{equation}
and where $G_{\rm p}(\vec{r},\vec{r}^\prime,k,\beta)$ is the scalar periodic Green's function defined in \eqref{eq:scperGreens6} in Appendix \ref{sect:scalarperiodicG}.
The periodic electric Green's dyadic can furthermore be factorized as
$\vec{G}_{\rm ep}(\vec{r},\vec{r}^\prime,k,\beta)=\widetilde{\vec{G}}_{\rm e}(\vec{r},\vec{r}^\prime,k,\beta)\eu^{\iu\beta (z-z^\prime)}$, 
where $\widetilde{\vec{G}}_{\rm e}(\vec{r},\vec{r}^\prime,k,\beta)$ is periodic in $z-z^\prime$ with period $p$, and similarly for
the periodic magnetic Green's dyadic $\vec{G}_{\rm mp}(\vec{r},\vec{r}^\prime,k,\beta)$.

\subsection{Periodic Green's dyadic}
The spectral representation of the electric Green's dyadic is given by
\begin{equation}\label{eq:Gerepr1}
\vec{G}_{\rm e}(\vec{r},\vec{r}^\prime,k)=\frac{1}{2\pi}\int_{-\infty}^{\infty}\vec{G}_{\rm e}(\vec{\rho},\vec{\rho}^\prime,k,\alpha)\eu^{\iu\alpha (z-z^\prime)}\diff\alpha,
\end{equation}
where
\begin{equation}\label{eq:Gerepr1b}
\vec{G}_{\rm e}(\vec{\rho},\vec{\rho}^\prime,k,\alpha)=\vec{G}_{\rm e}^0(\vec{\rho},\vec{\rho}^\prime,k,\alpha)
-\frac{1}{k^2}\hat{\vec{\rho}}\hat{\vec{\rho}}\delta(\vec{\rho}-\vec{\rho}^\prime),
\end{equation}
and which are based on \eqref{eq:Gedef0} and \eqref{eq:Gedef1} in Appendix \ref{sect:modeexp}.
From the definition \eqref{eq:EMdyaddef} it follows that $\vec{G}_{\rm e}(\vec{r},\vec{r}^\prime,k)$ can also be represented as
\begin{equation}\label{eq:Gerepr2}
\vec{G}_{\rm e}(\vec{r},\vec{r}^\prime,k)=\frac{1}{2\pi}\int_{-\infty}^{\infty}\{\vec{I}+\frac{1}{k^2}\nabla\nabla\}
G(\vec{\rho},\vec{\rho}^\prime,k,\alpha)\eu^{\iu\alpha(z-z^\prime)}\diff\alpha,
\end{equation}
where $G(\vec{\rho},\vec{\rho}^\prime,k,\alpha)$ is the scalar two-dimensional Green's function defined in \eqref{eq:scperGreens2}
in Appendix \ref{sect:scalarperiodicG}. A comparison of \eqref{eq:Gerepr1} and  \eqref{eq:Gerepr2} yields immediately that
\begin{multline}
\vec{G}_{\rm e}(\vec{\rho},\vec{\rho}^\prime,k,\alpha)\eu^{\iu\alpha (z-z^\prime)}\\
=\{\vec{I}+\frac{1}{k^2}\nabla\nabla\}G(\vec{\rho},\vec{\rho}^\prime,k,\alpha)\eu^{\iu\alpha(z-z^\prime)},
\end{multline}
and where both sides can be extended analytically into the complex $\alpha$-plane.
Based on the Poisson summation formula for the scalar periodic Green's function given in \eqref{eq:Gpexpr}, 
an analytic expression for the periodic electric Green's dyadic can now be derived as follows
\begin{multline}
\vec{G}_{\rm ep}(\vec{r},\vec{r}^\prime,k,\beta)=\{\vec{I}+\frac{1}{k^2}\nabla\nabla\} G_{\rm p}(\vec{r},\vec{r}^\prime,k,\beta)\\
=\frac{1}{p}\sum_{n=-\infty}^{\infty}\{\vec{I}+\frac{1}{k^2}\nabla\nabla\} G(\vec{\rho},\vec{\rho}^\prime,k,\beta+n\frac{2\pi}{p})\eu^{\iu(\beta+n\frac{2\pi}{p})(z-z^\prime)}\\
=\frac{1}{p}\sum_{n=-\infty}^{\infty} \vec{G}_{\rm e}(\vec{\rho},\vec{\rho}^\prime,k,\beta+n\frac{2\pi}{p})\eu^{\iu(\beta+n\frac{2\pi}{p})(z-z^\prime)}.
\end{multline}
Finally, based on \eqref{eq:Gerepr1b} the periodic electric Green's dyadic is given by
\begin{multline}\label{eq:GepGep0distribut}
\vec{G}_{\rm ep}(\vec{r},\vec{r}^\prime,k,\beta)
=\vec{G}_{\rm ep}^0(\vec{r},\vec{r}^\prime,k,\beta) \\
-\frac{1}{k^2}\hat{\vec{\rho}}\hat{\vec{\rho}}\delta(\vec{\rho}-\vec{\rho}^\prime)\sum_{n=-\infty}^{\infty}\delta(z-z^\prime+np)\eu^{\iu\beta (z-z^\prime)},
\end{multline}
where
\begin{multline}\label{eq:Gep0def}
\vec{G}_{\rm ep}^0(\vec{r},\vec{r}^\prime,k,\beta)\\
=\frac{1}{p}\sum_{n=-\infty}^{\infty} \vec{G}_{\rm e}^0(\vec{\rho},\vec{\rho}^\prime,k,\beta+n\frac{2\pi}{p})\eu^{\iu(\beta+n\frac{2\pi}{p})(z-z^\prime)}.
\end{multline}

Similarly, the following spectral representation of the magnetic Green's dyadic follows from \eqref{eq:EMdyaddef}
\begin{equation}\label{eq:Gmrepr2}
\vec{G}_{\rm m}(\vec{r},\vec{r}^\prime,k)=\frac{1}{2\pi}\int_{-\infty}^{\infty}
\nabla G(\vec{\rho},\vec{\rho}^\prime,k,\alpha)\eu^{\iu\alpha(z-z^\prime)}\times\vec{I} \diff\alpha,
\end{equation}
and hence 
\begin{equation}
\vec{G}_{\rm m}(\vec{\rho},\vec{\rho}^\prime,k,\alpha)\eu^{\iu\alpha (z-z^\prime)}
=\nabla G(\vec{\rho},\vec{\rho}^\prime,k,\alpha)\eu^{\iu\alpha(z-z^\prime)}\times\vec{I}.
\end{equation}
An analytic expression for the periodic magnetic Green's dyadic can then be derived as follows
\begin{multline}\label{eq:Gmpdef}
\vec{G}_{\rm mp}(\vec{r},\vec{r}^\prime,k,\beta)=\nabla G_{\rm p}(\vec{r},\vec{r}^\prime,k,\beta)\times\vec{I}\\
=\frac{1}{p}\sum_{n=-\infty}^{\infty}\nabla G(\vec{\rho},\vec{\rho}^\prime,k,\beta+n\frac{2\pi}{p}) \eu^{\iu(\beta+n\frac{2\pi}{p})(z-z^\prime)}\times\vec{I}\\
=\frac{1}{p}\sum_{n=-\infty}^{\infty} \vec{G}_{\rm m}(\vec{\rho},\vec{\rho}^\prime,k,\beta+n\frac{2\pi}{p})\eu^{\iu(\beta+n\frac{2\pi}{p})(z-z^\prime)}.
\end{multline}

\subsection{Integral equation for natural modes}
Consider now the source-free Maxwell's equations as in \eqref{eq:Maxwell1} and \eqref{eq:Maxwell1eqsource} (with $\vec{M}_{\rm s}=\vec{J}_{\rm s}=\vec{0}$)
and with equivalent sources $\vec{M}(\vec{r})$ and $\vec{J}(\vec{r})$ defined as in \eqref{eq:eqsourcedef}, and where the periodic material dyadics are defined as in \eqref{eq:chidef}.
For simplicity, a non-magnetic material is assumed here with $\vec{\chi}_{\rm m}(\vec{r})=\vec{0}$ and $\vec{\chi}_{\rm e}(\vec{r})=\vec{\chi}(\vec{r})$.
The fields are assumed to have a Floquet wavenumber $\beta$ with $\Im\beta\geq 0$ and can
be expressed as $\vec{E}(\vec{r})=\widetilde{\vec{E}}(\vec{r})\eu^{\iu\beta z}$
with a $p$-periodic part given by
\begin{equation}\label{eq:Floquetmodedef2}
\widetilde{\vec{E}}(\vec{r})=\sum_{m=-\infty}^{\infty}\sum_{n=-\infty}^{\infty}\vec{E}_{mn}(\vec{\rho})\eu^{\iu m\phi}\eu^{\iu n\frac{2\pi}{p}z}.
\end{equation}
The general solution \eqref{eq:Maxwell2persol} then leads to the following integral equation for the electric field
\begin{multline}\label{eq:inteqper1}
\left[\vec{I}+\hat{\vec{\rho}}\hat{\vec{\rho}}\cdot\vec{\chi}(\vec{r})\right]\cdot\vec{E}(\vec{r}) \\
-k^2\int_{S}\int_{0}^{p}\vec{G}_{\rm ep}^0(\vec{r},\vec{r}^\prime,k,\beta)\cdot\vec{\chi}(\vec{r}^\prime)\cdot\vec{E}(\vec{r}^\prime)\diff S^\prime \diff z^\prime
=\vec{0}, 
\end{multline}
where the integration is over one unit cell $V_{\rm c}$ and  $\vec{r}\in V_{\rm c}$, and where the contribution from the source-point has been extracted by evaluating
the periodic distribution in \eqref{eq:GepGep0distribut}.
By writing $\vec{G}_{\rm ep}^0(\vec{r},\vec{r}^\prime,k,\beta)=\widetilde{\vec{G}}_{\rm e}^0(\vec{r},\vec{r}^\prime,k,\beta)\eu^{\iu\beta (z-z^\prime)}$, the 
integral equation \eqref{eq:inteqper1} can also be written
\begin{multline}\label{eq:inteqper2}
\left[\vec{I}+\hat{\vec{\rho}}\hat{\vec{\rho}}\cdot\vec{\chi}(\vec{r})\right]\cdot\widetilde{\vec{E}}(\vec{r}) \\
-k^2\int_{0}^{a}\int_{0}^{2\pi}\int_{0}^{p}\widetilde{\vec{G}}_{\rm e}^0(\vec{r},\vec{r}^\prime,k,\beta)\cdot\vec{\chi}(\vec{r}^\prime)
\cdot\widetilde{\vec{E}}(\vec{r}^\prime)\rho^\prime\diff \rho^\prime\diff\phi^\prime \diff z^\prime \\ =\vec{0}, 
\end{multline}
where $\vec{r}\in V_{\rm c}$. Here, $\widetilde{\vec{G}}_{\rm e}^0(\vec{r},\vec{r}^\prime,k,\beta)$ is defined by \eqref{eq:Gep0def} and \eqref{eq:Gedef2} and can hence be written as
\begin{multline}\label{eq:Gep0tildedef}
\widetilde{\vec{G}}_{\rm e}^0(\vec{r},\vec{r}^\prime,k,\beta)\\
=\frac{1}{p}\sum_{m=-\infty}^{\infty}\sum_{n=-\infty}^{\infty}\vec{a}_m(\vec{\rho},\vec{\rho}^\prime,k,\beta+n\frac{2\pi}{p})\eu^{\iu m(\phi-\phi^\prime)}\eu^{\iu n\frac{2\pi}{p}(z-z^\prime)}.
\end{multline}
By employing the convolution theorem for two-dimensional Fourier series, the integral equation \eqref{eq:inteqper2} can finally be written
in Fourier space exactly as in \eqref{eq:integraleqFinal}.


\section{Applications}

\subsection{Twist-modes for multi-conductor power cables}\label{sect:twistmodes}
\begin{figure}[htb]
\begin{picture}(50,100)
\put(100,0){\makebox(50,90){\includegraphics[width=8cm]{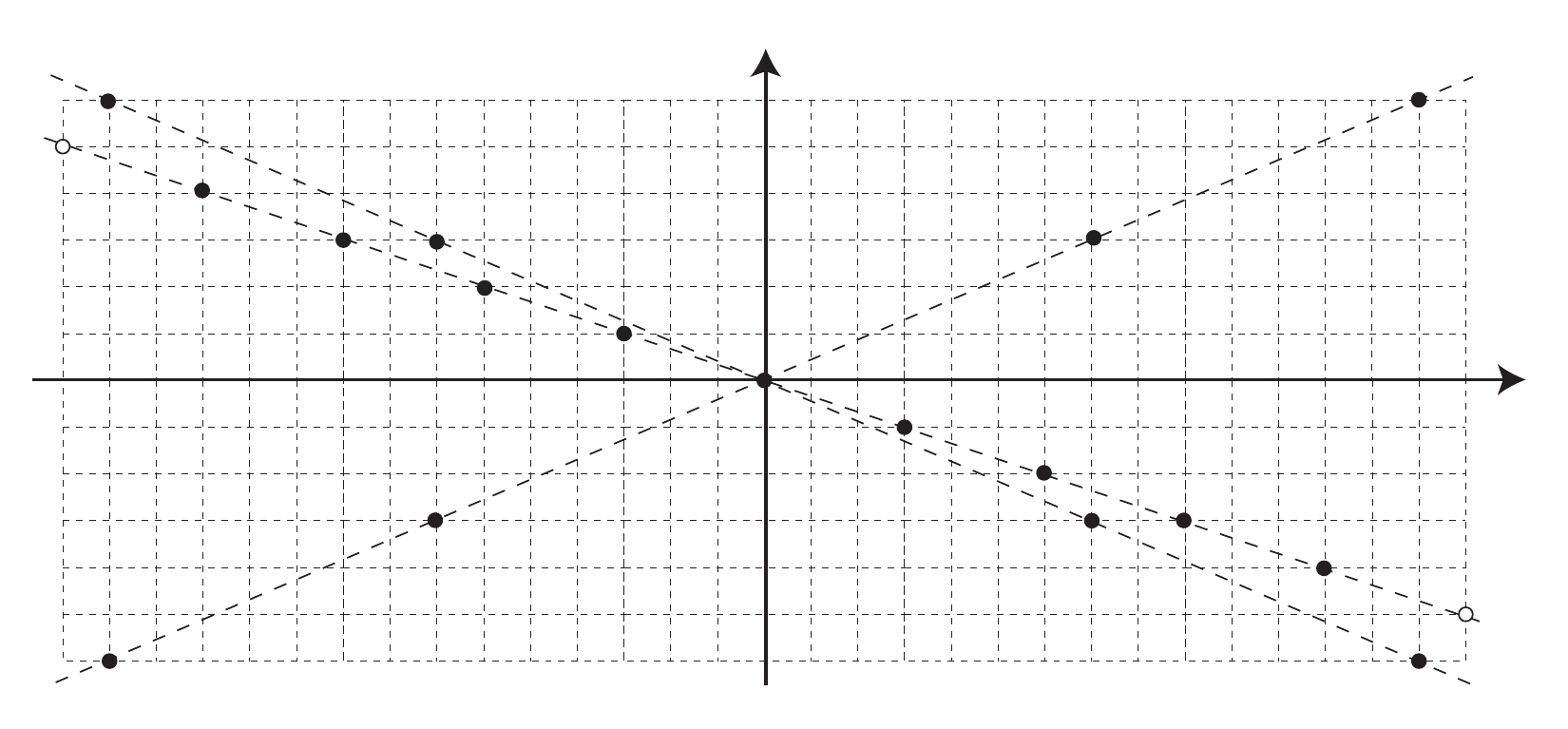}}} 
\put(230,48){\scriptsize $m$} 
\put(128,90){\scriptsize $n$} 
\put(227,7){\scriptsize ${\cal R}_1$} 
\put(227,-3){\scriptsize ${\cal R}_2$} 
\put(227,81){\scriptsize ${\cal R}_3$} 
\end{picture}
\caption{Material twist-modes $\vec{\chi}_{mn}$ in Fourier space. 
Here, ${\cal R}_1$, ${\cal R}_2$ and ${\cal R}_3$ denote three twist-regions with $(m_1,n_1)=(3,-1)$, $(m_2,n_2)=(7,-3)$ and $(m_3,n_3)=(7,3)$.
The filled bullets correspond to a possible truncation.}
\label{fig:helicalfig2}
\end{figure}
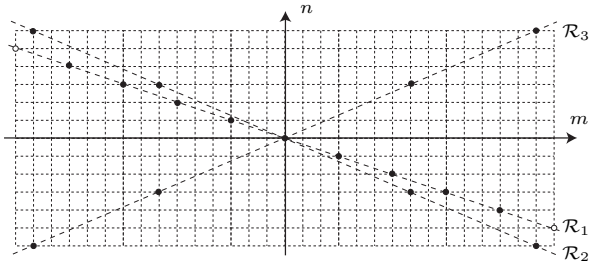

Assume that the cross-section of the waveguide rotates along the longitudinal direction and 
that there are $R$ radial regions ${\cal R}_r$ with distinct rotation (twist) for $r=1,\ldots,R$. 
In each region, the period in the azimuthal direction is given by $2\pi/m_r$ where $m_r\in\{1,2,\ldots\}$.
It is furthermore assumed that there is a smallest common period $p$ in the longitudinal direction such that the longitudinal period $p_r$ 
in each radial region is given by $p_r=p/n_r$ where $n_r\in\{\pm 1,\pm 2,\ldots\}$, and 
where the positive (negative) sign indicates a left (right) handed twist. The twist in each radial region ${\cal R}_r$ is thus 
characterized\footnote{Note that the associated twist angle $\varphi_r(\rho)=\arctan(\rho\frac{\diff \phi}{\diff z})=\arctan(\rho\frac{2\pi}{p}\frac{(-n_r)}{m_r})$
increases with the radius $\rho$.} by the twist direction $(m_r,n_r)$, and the non-zero Fourier coefficients $\vec{\chi}_{mn}(\vec{\rho})$ defined in
\eqref{eq:chidef} can only be found at the following points in the reciprocal (Fourier) space 
\begin{equation}\label{eq:Findicesthreephasechi}
\left\{\begin{array}{l}
m=km_r, \vspace{0.2cm} \\
n=kn_r,
\end{array}\right.
\end{equation}
where $k=0,\pm 1, \pm 2,\ldots$ and $r=1,\ldots,R$.
An example with three distinct (one left and two right handed) twist regions is illustrated in Fig.~\ref{fig:helicalfig2}.

The primary interest here is with the twisted modes of three-phase power cables at quasi-static (50\unit{Hz}) conditions.
Here, the excitation \eqref{eq:Fmrhoalpha} is governed by a single symmetric component \cite{Wedepohl1963} with 
positive phase progression, \ie an azimuthal Floquet-mode with factor $\eu^{\iu\phi}$. 
The corresponding field twist-mode $\vec{E}_{mn}(\vec{\rho})$ is defined by the excitation point $(m,n)=(1,0)$ 
together with the conditions of \eqref{eq:integraleqFinal}, and which yields the following set of feasible Fourier indices

\begin{equation}\label{eq:FindicesthreephaseE}
\left\{\begin{array}{l}
m=km_r+1, \vspace{0.2cm} \\
n=kn_r,
\end{array}\right.
\end{equation}
where $k=0,\pm 1, \pm 2,\ldots$ and $r=1,\ldots,R$. 
Note that this is simply a one-step shift to the right in comparison
to the illustration in Fig.~\ref{fig:helicalfig2}.

\subsection{Thin helical wires under quasi-static conditions}

A useful application of the periodic Green's dyadics expressed in \eqref{eq:Maxwell2persol} is with the calculation of the
electromagnetic fields produced by a helical current distribution. As an example, consider
the periodic magnetic field produced by a thin helical wire carrying the stationary current $I$ under quasi-static conditions. 
The radius of the helix is $a$ and the period is $p$. Here,
\begin{equation}\label{eq:Hhelix1}
\vec{H}(\vec{r})=I\int_{L}\vec{G}_{\rm mp}(\vec{r},\vec{r}^\prime,k,0)\cdot\diff{\vec{r}^\prime},
\end{equation}
where $\beta=0$ and $L$ is a curve constituting one period of the helix defined in cartesian coordinates $(x^\prime,y^\prime,z^\prime)$ as
\begin{equation}
\left\{\begin{array}{l}
x^\prime=a\cos(\frac{2\pi}{p}z^\prime+\varphi), \vspace{0.2cm} \\
y^\prime=a\sin(\frac{2\pi}{p}z^\prime+\varphi), \\
\end{array}\right.
\end{equation}
where $z^\prime\in[0,p]$ and $\varphi$ is an off-set parameter of the helix.
In cylindrical coordinates it is seen that $\rho^\prime=a$, $\phi^\prime=\frac{2\pi}{p}z^\prime+\varphi$
and $\frac{\diff \vec{r}^\prime}{\diff z^\prime}={\hat{\vec{\phi}}}^\prime a\frac{2\pi}{p}+\hat{\vec{z}}$.
By using \eqref{eq:Gmpdef}, \eqref{eq:Gmdef2} and \eqref{eq:dyadb} for $\rho>a$,
the expression \eqref{eq:Hhelix1} yields the result
\begin{multline}\label{eq:Hhelix2}
\vec{H}(\vec{r})=I\frac{\iu k}{4}\sum_{m=-\infty}^{\infty}\sum_{\tau=1}^2
\vec{u}_{\bar{\tau} m}(\vec{\rho},-m\frac{2\pi}{p})\vec{v}_{\tau m}^\dagger(\vec{\rho}^{\prime},-m\frac{2\pi}{p})\\
\cdot({\hat{\vec{\phi}}}^\prime a\frac{2\pi}{p}+\hat{\vec{z}})\eu^{\iu m(\phi-\varphi-\frac{2\pi}{p}z)},
\end{multline}
where the integration over one period has been performed as
\begin{equation}
\int_{0}^{p}\eu^{-\iu(m+n)\frac{2\pi}{p}z^\prime}\diff z^\prime=p \delta_{-m,n}.
\end{equation}


\section{Numerical method}

\subsection{Discretization by the collocation method}
A discretization based on the collocation method \cite{Kress1999} is devised as follows.
It is assumed that the material region of the helical structure is given by the radial domain $\rho_1\leq\rho\leq a$ where $\rho_1>0$.
An $N$-point discretization is defined where $\rho_1<\rho_2<\ldots<\rho_N=a$ and $L_j(\rho)$ denotes the corresponding Lagrange basis consisting
of linear splines with the interpolation property $L_j(\rho_i)=\delta_{ij}$ for $i,j=1,\ldots,N$, see \eg \cite{Kress1999}.
The Fourier components of the electric field defined in \eqref{eq:Floquetmodedef} or \eqref{eq:Floquetmodedef2} is now expanded as
\begin{equation}\label{eq:EmnLj}
\vec{E}_{mn}(\vec{\rho})=\sum_{j=1}^N\vec{E}_{mnj}(\vec{\rho})L_j(\rho),
\end{equation}
where $\vec{E}_{mnj}(\vec{\rho})$ is a vector valued coefficient with constant cylindrical components.
Let $\vec{\rho}_i$ denote the radial vector $\vec{\rho}$ evaluated at the point $\rho_i$ for $i=1,\ldots,N$.
Due to the interpolation property of the Lagrange basis it is seen that $\vec{E}_{mn}(\vec{\rho}_i)=\vec{E}_{mni}(\vec{\rho})$.
The integral equation \eqref{eq:integraleqFinal} evaluated at the interpolation points $\rho_i$ now yields the discrete system
\begin{multline}\label{eq:integral4num}
\vec{E}_{mni}(\vec{\rho})+\sum_{m^\prime}\sum_{n^\prime}
\hat{\vec{\rho}}\hat{\vec{\rho}}\cdot\vec{\chi}_{m-m^\prime,n-n^\prime}(\vec{\rho}_i)\cdot\vec{E}_{m^\prime n^\prime i}(\vec{\rho})\\
-k^22\pi\sum_{m^\prime}\sum_{n^\prime}\sum_{j=1}^N\int_{\Omega_j}\vec{a}_m(\vec{\rho}_i,\vec{\rho}^\prime,\beta+n\frac{2\pi}{p})\\
\cdot\vec{\chi}_{m-m^\prime,n-n^\prime}(\vec{\rho}^\prime)L_j(\rho^\prime)\cdot\vec{E}_{m^\prime n^\prime j}(\vec{\rho}^\prime)\rho^\prime\diff\rho^\prime=\vec{0},
\end{multline}
where $\Omega_j$ denotes the support region of $L_j(\rho)$.

The Fourier series \eqref{eq:Floquetmodedef2} is truncated using $M$ terms based on the assumed twist-modes that are at hand, \cf section \ref{sect:twistmodes}.
A multi-index notation is introduced where $k\leftrightarrow(m,n)$, $l\leftrightarrow(m^\prime,n^\prime)$ and where $k,l=1,\ldots,M$.
The following definitions are made 
\begin{equation}\label{eq:Eki}
\vec{E}_{ki}=\vec{E}_{mni}(\vec{\rho}), \quad \quad   \vec{E}_{lj}=\vec{E}_{m^\prime n^\prime j}(\vec{\rho}),
\end{equation}
\begin{equation}\label{eq:chikil}
\vec{\chi}_{kil}=\vec{\chi}_{m-m^\prime,n-n^\prime}(\vec{\rho}_i),
\end{equation}
and
\begin{multline}\label{eq:akilj}
\vec{a}_{kilj}(\beta)=
\int_{\Omega_j}\vec{a}_m(\vec{\rho}_i,\vec{\rho}^\prime,\beta+n\frac{2\pi}{p})\\
\cdot\vec{\chi}_{m-m^\prime,n-n^\prime}(\vec{\rho}^\prime)L_j(\rho^\prime)\rho^\prime\diff\rho^\prime,
\end{multline}
where $k,l=1\ldots,M$ and $i,j=1,\ldots,N$. The integration in \eqref{eq:akilj} is highly regular with at most some points of discontinuity in either of the
terms $\vec{a}_m(\vec{\rho}_i,\vec{\rho}^\prime,\beta+n\frac{2\pi}{p})$ and $\vec{\chi}_{m-m^\prime,n-n^\prime}(\vec{\rho}^\prime)$.
It is therefore convenient to place any possible points of discontinuity of the material function $\vec{\chi}_{mn}(\vec{\rho}^\prime)$ 
at the grid points $\rho_i$ of the linear interpolation, and to employ an efficient quadrature rule based on
interior points such as the Gauss-Legendre quadrature \cite{Kress1999,Olver+etal2010} to evaluate the integral in \eqref{eq:akilj}.
The system \eqref{eq:integral4num} can now be written in the more convenient form
\begin{equation}\label{eq:integral5num}
\vec{E}_{ki}+\sum_{l=1}^M\hat{\vec{\rho}}\hat{\vec{\rho}}\cdot\vec{\chi}_{kil}\cdot\vec{E}_{li}
-k^22\pi\sum_{l=1}^M\sum_{j=1}^N\vec{a}_{kilj}(\beta)\cdot\vec{E}_{lj}=\vec{0},
\end{equation}
which can be readily interpreted in terms of $3\times 3$ block matrices and with the row and column multi-indices $(k,i)$ and $(l,j)$, respectively.

Let ${\bf A}(\beta)$ denote the finite matrix corresponding to the linear system defined in \eqref{eq:integral5num}.
The propagation constant can then be computed from a numerical residue calculus based on
\begin{equation}\label{eq:contoureig}
\beta_1=\displaystyle\frac{\displaystyle\oint_{{\cal C}}\displaystyle\frac{\beta}{\lambda_1\{{\bf A}(\beta)\}}\diff\beta}
{\displaystyle\oint_{{\cal C}}\displaystyle\frac{1}{\lambda_1\{{\bf A}(\beta)\}}\diff\beta},
\end{equation}
where $\lambda_1\{{\bf A}(\beta)\}$ is the minimum (modulus) eigenvalue of the matrix ${\bf A}(\beta)$, and where
the closed loop ${\cal C}$ is circumscribing the true zero $\beta_{1}$ in such a way that there are no other zeros 
or branch-points of $\lambda_1\{{\bf A}(\beta)\}$ inside the loop.

\subsection{Material smoothing}

\begin{figure}[htb]
\begin{center}
\scriptsize
\input{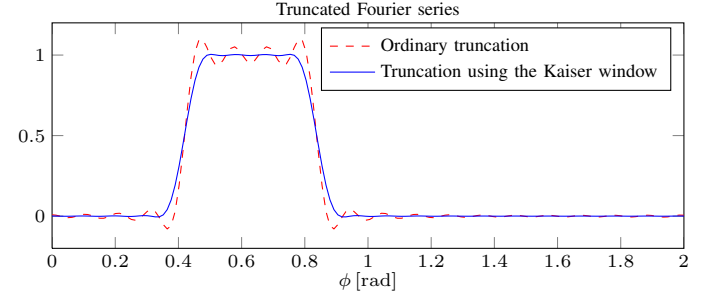}
\end{center}
\vspace{-5mm}
\caption{Truncation of the Fourier series of a square pulse, with and without spectral smooting and with azimuthal indices $m=-20,\ldots,20$.}
\label{fig:matfig1}
\end{figure}

Appropriate windowing techniques can be applied directly in the Fourier domain to reduce the Gibbs phenomena in the truncation of
the (material) Fourier coefficients $\vec{\chi}_{mn}(\vec{\rho})$. As an example is illustrated in Fig.~\ref{fig:matfig1} a comparison of the truncated 
Fourier series expansions of a square pulse with and without a one-dimensional coefficient smoothing 
based on the Kaiser window \cite{Oppenheim+Schafer1999}. 

%
%

\appendices
\section{Mode expansions of the dyadic Green's functions}\label{sect:modeexp}
Consider a homogeneous and isotropic cylindrical region with relative permittivity $\epsilon$, relative permeability $\mu$ and
wavenumber $k=k_0\sqrt{\mu\epsilon}$.
The solenoidal (source-free) cylindrical vector waves are defined here by
\begin{equation}\label{eq:cylvecdef}
\left\{\begin{array}{l}
\vec{w}_{1m}(\vec{r},\alpha)=\displaystyle\frac{1}{\kappa}\nabla\times\left(\hat{\vec{z}}\psi_m(\kappa\rho)\eu^{\iu m\phi}\eu^{\iu\alpha z}\right), \vspace{0.2cm} \\
\vec{w}_{2m}(\vec{r},\alpha)=\displaystyle\frac{1}{k}\nabla\times\vec{w}_{1m}(\vec{r}),
\end{array}\right.
\end{equation}
where $\psi_m(\kappa\rho)$ is a Bessel function or a Hankel function of the first kind, each of order $m$, \cf \cite{Bostrom+Kristensson+Strom1991,Collin1991}.
Here, $\alpha$ is the longitudinal wavenumber and $\kappa=\sqrt{k^2-\alpha^2}$ the transversal wavenumber where
the square root is chosen such that $0<\arg\kappa\leq \pi$ and hence $\Im \kappa\geq 0$. 
It can be shown by direct calculation that $\nabla\times \vec{w}_{2m}(\vec{r},\alpha)=k\vec{w}_{1m}(\vec{r},\alpha)$.
The following curl properties are thus obtained
\begin{equation}\label{eq:w12cross}
\nabla\times \vec{w}_{\tau m}(\vec{r},\alpha)=k\vec{w}_{\bar{\tau} m}(\vec{r},\alpha),
\end{equation}
for  $\tau=1,2$, and where $\bar{\tau}$ denotes the complement of $\tau$ ($\bar{1}=2$ and $\bar{2}=1$).

The following notation will be used 
\begin{equation}\label{eq:cylvecexpr2}
\vec{w}_{\tau m}(\vec{r},\alpha)=\vec{w}_{\tau m}(\vec{\rho},\alpha)\eu^{\iu m\phi}\eu^{\iu\alpha z},
\end{equation}
where $\tau=1,2$, and the vectors $\vec{w}_{\tau m}(\vec{\rho},\alpha)$ are given explicitly 
in cylindrical coordinates as
\begin{equation}\label{eq:cylvecexpr}
\left\{\begin{array}{l}
\vec{w}_{1m}(\vec{\rho},\alpha)=\displaystyle \hat{\vec{\rho}}\frac{\iu m}{\kappa\rho}\psi_{m}(\kappa\rho)
-\hat{\vec{\phi}}\psi_{m}^{\prime}(\kappa\rho),  \vspace{0.2cm} \\
\vec{w}_{2m}(\vec{\rho},\alpha)=\displaystyle \hat{\vec{\rho}}\frac{\iu\alpha}{k}\psi_{m}^{\prime}(\kappa\rho)
-\hat{\vec{\phi}}\frac{m\alpha}{k\kappa\rho}\psi_{m}(\kappa\rho)+\hat{\vec{z}}\frac{\kappa}{k}\psi_{m}(\kappa\rho),
\end{array}\right.
\end{equation}
and where the $^\prime$ denotes differentiation with respect to the argument.
Let the regular and the outgoing (radiating) cylindrical vector waves $\vec{v}_{\tau m}(\vec{r},\alpha)$ and $\vec{u}_{\tau m}(\vec{r},\alpha)$ be defined 
as in (\ref{eq:cylvecdef}) by using the regular Bessel functions and the Hankel functions of the first kind, ${\rm J}_m(\kappa\rho)$ and ${\rm H}_m^{(1)}(\kappa\rho)$, 
respectively. 

The electric dyadic Green's function defined in \eqref{eq:EMdyaddef} can be expanded in cylindrical vector waves as  
\begin{multline}\label{eq:Greensdyadic2}
\vec{G}_{\rm e}(\vec{r},\vec{r}^\prime,k) \\
=\frac{\iu}{8\pi}\int_{-\infty}^{\infty}\sum_{m=-\infty}^{\infty}\sum_{\tau=1}^2
\vec{v}_{\tau m}(\vec{r}_{<},\alpha)\vec{u}_{\tau m}^\dagger(\vec{r}_{>},\alpha)\diff\alpha \\
-\frac{1}{k^2}\hat{\vec{\rho}}\hat{\vec{\rho}}\delta(\vec{r}-\vec{r}^\prime),
\end{multline}
where $\vec{r}_{<}$  and $\vec{r}_{>}$ denote the vector in $\{\vec{r},\vec{r}^\prime\}$ having the smallest and largest radial coordinate,
respectively, \ie $\rho_{<}=\min\{\rho,\rho^\prime\}$ and $\rho_{>}=\max\{\rho,\rho^\prime\}$, \cf \cite{Bostrom+Kristensson+Strom1991,Collin1991,Chew1995}.
The dagger $\dagger$ refers to a sign-shift in the exponentials in the definition \eqref{eq:cylvecdef},
and which can be placed on any of the two vector waves $\vec{v}_{\tau m}(\vec{r},\alpha)$ and $\vec{u}_{\tau m}(\vec{r},\alpha)$ 
in the expression \eqref{eq:Greensdyadic2}, \cf \cite{Bostrom+Kristensson+Strom1991}.
Note that the expression \eqref{eq:Greensdyadic2} also contains a delta distribution taking the source point into account, \cf \cite{Collin1991}.

By employing the notation defined in \eqref{eq:cylvecexpr2}, the electric Green's dyadic can now be expressed as
\begin{equation}\label{eq:Gedef0}
\vec{G}_{\rm e}(\vec{r},\vec{r}^\prime,k)=\vec{G}_{\rm e}^0(\vec{r},\vec{r}^\prime,k)-\frac{1}{k^2}\hat{\vec{\rho}}\hat{\vec{\rho}}\delta(\vec{r}-\vec{r}^\prime)
\end{equation}
where
\begin{equation}\label{eq:Gedef1}
\vec{G}_{\rm e}^0(\vec{r},\vec{r}^\prime,k)=\frac{1}{2\pi}\int_{-\infty}^{\infty}\vec{G}_{\rm e}^0(\vec{\rho},\vec{\rho}^\prime,k,\alpha)\eu^{\iu\alpha (z-z^\prime)}\diff\alpha,
\end{equation}
\begin{equation}\label{eq:Gedef2}
\vec{G}_{\rm e}^0(\vec{\rho},\vec{\rho}^\prime,k,\alpha)=\sum_{m=-\infty}^{\infty}\vec{a}_m(\vec{\rho},\vec{\rho}^\prime,k,\alpha)\eu^{\iu m(\phi-\phi^\prime)},
\end{equation}
and where the dyadic $\vec{a}_m(\vec{\rho},\vec{\rho}^\prime,k,\alpha)$ is given by
\begin{equation}\label{eq:dyada}
\vec{a}_m(\vec{\rho},\vec{\rho}^\prime,k,\alpha)=\left\{
\begin{array}{ll}
\displaystyle\frac{\iu}{4}\sum_{\tau=1}^2\vec{u}_{\tau m}(\vec{\rho},\alpha)\vec{v}_{\tau m}^\dagger(\vec{\rho}^{\prime},\alpha) & \rho^\prime <\rho, \vspace{0.2cm} \\
\displaystyle\frac{\iu}{4}\sum_{\tau=1}^2\vec{v}_{\tau m}(\vec{\rho},\alpha)\vec{u}_{\tau m}^\dagger(\vec{\rho}^{\prime},\alpha) & \rho^\prime >\rho,
\end{array}\right.
\end{equation}
where the dagger $\dagger$ has been placed on the vector with primed coordinates, and where 
$\vec{w}_{\tau m}^{\dagger}(\vec{r}^\prime,\alpha)=\vec{w}_{\tau m}^{\dagger}(\vec{\rho}^\prime,\alpha)\eu^{-\iu m\phi^\prime}\eu^{-\iu\alpha z^\prime}$.

The magnetic dyadic Green's function can be obtained as $\vec{G}_{\rm m}(\vec{r},\vec{r}^\prime,k)=\nabla\times\vec{G}_{\rm e}(\vec{r},\vec{r}^\prime,k)$ \cite{Jin2010},
and it follows from \eqref{eq:w12cross} and \eqref{eq:Greensdyadic2} that it can be expanded in cylindrical vector waves as  
\begin{equation}\label{eq:Gmdef1}
\vec{G}_{\rm m}(\vec{r},\vec{r}^\prime,k)=\frac{1}{2\pi}\int_{-\infty}^{\infty}\vec{G}_{\rm m}(\vec{\rho},\vec{\rho}^\prime,k,\alpha)\eu^{\iu\alpha (z-z^\prime)}\diff\alpha,
\end{equation}
where
\begin{equation}\label{eq:Gmdef2}
\vec{G}_{\rm m}(\vec{\rho},\vec{\rho}^\prime,k,\alpha)=\sum_{m=-\infty}^{\infty}\vec{b}_m(\vec{\rho},\vec{\rho}^\prime,k,\alpha)\eu^{\iu m(\phi-\phi^\prime)},
\end{equation}
and where the dyadic $\vec{b}_m(\vec{\rho},\vec{\rho}^\prime,k,\alpha)$ is given by
\begin{equation}\label{eq:dyadb}
\vec{b}_m(\vec{\rho},\vec{\rho}^\prime,k,\alpha)=\left\{
\begin{array}{ll}
\displaystyle\frac{\iu k}{4}\sum_{\tau=1}^2\vec{u}_{\bar{\tau} m}(\vec{\rho},\alpha)\vec{v}_{\tau m}^\dagger(\vec{\rho}^{\prime},\alpha) & \rho^\prime <\rho, \vspace{0.2cm} \\
\displaystyle\frac{\iu k}{4}\sum_{\tau=1}^2\vec{v}_{\bar{\tau} m}(\vec{\rho},\alpha)\vec{u}_{\tau m}^\dagger(\vec{\rho}^{\prime},\alpha) & \rho^\prime >\rho.
\end{array}\right.
\end{equation}

\section{Scalar periodic Green's function}\label{sect:scalarperiodicG}
The scalar Green's function $G(\vec{r},\vec{r}^\prime,k)=\frac{\eu^{\iu k|\vec{r}-\vec{r}^\prime|}}{4\pi|\vec{r}-\vec{r}^\prime|}$ satisfies the 
inhomogeneous Helmholtz equation
\begin{equation}\label{eq:scperGreens1}
\left\{\nabla^2+k^2 \right\}G(\vec{r},\vec{r}^\prime,k)=-\delta(\vec{r}-\vec{r}^\prime),
\end{equation}
and can be represented by the Fourier integral
\begin{equation}\label{eq:scperGreens2}
G(\vec{r},\vec{r}^\prime,k)=\frac{1}{2\pi}\int_{-\infty}^{\infty}G(\vec{\rho},\vec{\rho}^\prime,k,\alpha)\eu^{\iu\alpha (z-z^\prime)}\diff\alpha,
\end{equation}
where the two-dimensional Green's function $G(\vec{\rho},\vec{\rho}^\prime,k,\alpha)$ satisfies
\begin{equation}\label{eq:scperGreens3}
\left\{\nabla_{\rm t}^2+\kappa^2 \right\}G(\vec{\rho},\vec{\rho}^\prime,k,\alpha)=-\delta(\vec{\rho}-\vec{\rho}^\prime),
\end{equation}
and where $\nabla_{\rm t}^2$ is the two-dimensional transverse Laplace operator, $\kappa=\sqrt{k^2-\alpha^2}$ ($\Im\kappa\geq 0$) and $\delta(\vec{\rho}-\vec{\rho}^\prime)$ the transverse
delta distribution. The function $G(\vec{\rho},\vec{\rho}^\prime,k,\alpha)$ is given by
\begin{equation}\label{eq:scperGreens4}
G(\vec{\rho},\vec{\rho}^\prime,k,\alpha)=\frac{\iu}{4}{\rm H}_{0}^{(1)}(\kappa|\vec{\rho}-\vec{\rho}^\prime|),
\end{equation}
and which can be expanded as
\begin{equation}\label{eq:scperGreens5}
G(\vec{\rho},\vec{\rho}^\prime,k,\alpha)=\frac{\iu}{4}\sum_{m=-\infty}^{\infty}{\rm J}_{m}(\kappa\rho_{<}){\rm H}_{m}^{(1)}(\kappa\rho_{>})\eu^{\iu m(\phi-\phi^\prime)},
\end{equation}
see \eg \cite{Bostrom+Kristensson+Strom1991,Peterson+Ray+Mittra1998,Jin2010} and the Graf's and Gegenbauer's addition theorem \cite{Olver+etal2010}.

The periodic Green's function $G_{\rm p}(\vec{r},\vec{r}^\prime,k,\beta)$ 
for plane wave excitation $\eu^{\iu\beta z}$ satisfies
\begin{multline}\label{eq:scperGreens6}
\left\{\nabla^2+k^2 \right\}G_{\rm p}(\vec{r},\vec{r}^\prime,k,\beta)\\
=-\delta(\vec{\rho}-\vec{\rho}^\prime)\sum_{n=-\infty}^{\infty}\delta(z-z^\prime+np)\eu^{\iu\beta (z-z^\prime)},
\end{multline}
together with a radiation condition in the transverse plane, and where $p$ is the period and $\Im\beta\geq 0$.
The periodic Green's function can furthermore be factorized as
\begin{equation}\label{eq:scperGreens7}
G_{\rm p}(\vec{r},\vec{r}^\prime,k,\beta)=\widetilde{G}(\vec{r},\vec{r}^\prime,k,\beta)\eu^{\iu\beta (z-z^\prime)},
\end{equation}
where $\widetilde{G}(\vec{r},\vec{r}^\prime,k,\beta)$ is periodic in $z-z^\prime$ with period $p$.

To derive an analytic expression for the periodic Green's function \cite{Peterson+Ray+Mittra1998}, the following 
Poisson summation formula \cite{Grafakos2008} can be employed 
\begin{equation}\label{eq:poisson1}
\sum_{n=-\infty}^{\infty}f(np)\eu^{\iu\beta pn}=\frac{1}{p}\sum_{n=-\infty}^{\infty}\hat{F}(\beta+n\frac{2\pi}{p}),
\end{equation}
where $\beta$ is real valued and where the Fourier transform is defined by
\begin{equation}\label{eq:poisson2}
\left\{\begin{array}{l}
\hat{F}(\alpha)=\displaystyle\int_{-\infty}^{\infty}f(z)\eu^{\iu\alpha z}\diff z, \vspace{0.2cm} \\
f(z)=\displaystyle\frac{1}{2\pi}\int_{-\infty}^{\infty}\hat{F}(\alpha)\eu^{-\iu\alpha z}\diff\alpha.
\end{array}\right.
\end{equation}
The periodic samples of $G(\vec{r},\vec{r}^\prime,k)$ along the $z^\prime$-coordinate are given by
\begin{equation}\label{eq:scperGreens8}
G(\vec{r},\vec{r}^\prime+\hat{\vec{z}}np,k)=\frac{1}{2\pi}\int_{-\infty}^{\infty}G(\vec{\rho},\vec{\rho}^\prime,k,\alpha)\eu^{\iu\alpha (z-z^\prime)}\eu^{-\iu\alpha np}\diff\alpha,
\end{equation}
and it follows from the Poisson summation formula \eqref{eq:poisson1} that
\begin{multline}\label{eq:Gpexpr}
G_{\rm p}(\vec{r},\vec{r}^\prime,k,\beta)=\sum_{n=-\infty}^{\infty}G(\vec{r},\vec{r}^\prime+\hat{\vec{z}}np,k)\eu^{\iu\beta pn} \\
=\frac{1}{p}\sum_{n=-\infty}^{\infty}G(\vec{\rho},\vec{\rho}^\prime,k,\beta+n\frac{2\pi}{p})\eu^{\iu(\beta+n\frac{2\pi}{p}) (z-z^\prime)}.
\end{multline}
It is noted that the first sum to the left in \eqref{eq:Gpexpr} converges only conditionally (not absolutely) when $\beta$ is real valued and it diverges
when $\Im\beta\neq 0$, whereas the sum on the right-hand side of \eqref{eq:Gpexpr} converges exponentially for all $\beta$ with $\Im\beta\geq 0$.
By using the property \eqref{eq:scperGreens3} which is valid with complex valued wavenumbers $\kappa$, it is readily seen that the expression in the
right-hand side of \eqref{eq:Gpexpr} satisfies \eqref{eq:scperGreens6} when $\beta$ is complex valued.

\begin{IEEEbiographynophoto}{Sven~Nordebo}
received the M.S.\ degree in electrical engineering from the Royal Institute of Technology, Stockholm, Sweden, in 1989, 
and the Ph.D.\ degree in signal processing from Lule{\aa} University of Technology, Lule{\aa}, Sweden, in 1995. 
He was appointed Docent in signal processing 1999. Since 2002 he is a Professor of Signal Processing at the 
Department of Physics and Electrical Engineering at the Linnaeus University. During 2009-2012 he was a Guest Professor 
of Signal Processing at the Department of Electrical and Information Technology, Lund University.
His research interests are in statistical signal processing, estimation theory, electromagnetics, 
antennas and propagation, inverse problems and imaging, optimization, dispersion modeling and analysis,
direct and inverse problems in applied waveguide theory.
\end{IEEEbiographynophoto}\vspace{-0.5cm}


\begin{IEEEbiographynophoto}{Mats~Gustafsson}
received the M.Sc. degree in Engineering Physics 1994, the Ph.D. degree in Electromagnetic Theory 2000, 
was appointed Docent 2005, and Professor of Electromagnetic Theory 2011, all from Lund University, Sweden. 
He co-founded the company Phase holographic imaging AB in 2004.  His research interests are in scattering 
and antenna theory and inverse scattering and imaging with applications in microwave tomography and digital holography. 
He has written over 65 peer reviewed journal papers and over 85 conference papers. 
Prof. Gustafsson received the Best Antenna Poster Prize at EuCAP 2007, the IEEE Schelkunoff Transactions 
Prize Paper Award 2010, and the Best Antenna Theory Paper Award at EuCAP 2013. 
He serves as an IEEE AP-S Distinguished Lecturer for 2013-15.
\end{IEEEbiographynophoto}\vspace{-0.5cm}


\begin{IEEEbiographynophoto}{Gerhard Kristensson}
was born 1949, he received his B.S. degree in mathematics and physics in 1973, and the Ph.D. degree in theoretical physics in 1979, 
both from the University of G\"{o}teborg, Sweden. In 1983 he was appointed Docent in theoretical physics at the University of G\"{o}teborg.
During 1977--1984 he held a research position sponsored by the National Swedish Board for Technical Development (STU) and he was Lecturer 
at the Institute of Theoretical Physics, G\"{o}teborg from 1980--1984.
In 1984--1986 he was a Visiting Scientist at the Applied Mathematical Sciences group, Ames Laboratory, Iowa State University.
He held a Docent position at the Department of Electromagnetic Theory, Royal Institute of Technology, Stockholm during 1986--1989, 
and in 1989 he was appointed the Chair of Electromagnetic Theory at Lund Institute of Technology, Sweden.
%
%
Kristensson is the author of 4 textbooks and the editor of 3 scientific books. He has written 12 chapters in scientific books, and he is the author of 
over 70 peer reviewed journal papers and over 70 reviewed contributions in conference proceedings.
Kristensson's major research interests are focused on wave propagation in inhomogeneous media, especially inverse scattering problems.
High frequency scattering methods, asymptotic expansions, optical fibers, antenna problems, and mixture formulas are also of interest, 
as well as radome design problems and homogenization of complex materials.
\end{IEEEbiographynophoto}\vspace{-0.5cm}

\begin{IEEEbiographynophoto}{B\"{o}rje Nilsson}
received the M.S. degree in mathematics and physics from G\"oteborg University, Sweden, in 1971, and the Ph.D. degree
in theoretical physics from the same university in 1980. He has 13 years industrial experience in acoustics and has been visiting professor in engineering acoustics from 1988 to 1994 at the Royal Institute of Technology, Stockholm, Sweden. He was appointed Docent in Engineering acoustics 1994. Since 2005 he is a Professor of Mathematical Physics at the Department of Mathematics at the Linn\ae us University. His main research interest is mathematical modelling of wave phenomena. Primarily electromagnetic, acoustic and quantum waves are considered. Of special interest are inverse problems and scattering theory in waveguide systems.
\end{IEEEbiographynophoto}\vspace{-0.5cm}


\begin{IEEEbiographynophoto}{Alexander I. Nosich}
(M'94-SM'95-F'04) was born in Kharkiv, Ukraine, in 1953. He received his M.S., Ph.D., and D.Sc. degrees in Radio Physics from the Kharkiv National University in 1975, 1979, and 1990, respectively. Since 1979, he has been with the Institute of Radio Physics and Electronics of the National Academy of Science of Ukraine, Kharkov, where he is currently a professor and principal scientist. Since 2010, he also heads initiated by him Laboratory of Micro and Nano Optics at this institute. Since 1992, he has held a number of guest fellowships and professorships in the EU, Japan, Singapore, and Turkey. His research interests include the method of analytical regularization, propagation and scattering of waves, simulation of lasers, open waveguides, and antennas, and the history of microwaves.
Prof. Nosich was one of the initiators of the international conference series on Mathematical Methods in Electromagnetic Theory (MMET) held in Ukraine in 1990-2014. In 1995, he organized IEEE East Ukraine Joint Chapter, the first in the former USSR. From 2001 to 2003, he represented Ukraine in the European Microwave Association. Since 2008, he represents Ukraine in the European Association on Antennas and Propagation.
\end{IEEEbiographynophoto}\vspace{-0.5cm}


\begin{IEEEbiographynophoto}{Daniel Sj\"{o}berg}
received the M.Sc. degree in engineering physics and the Ph.D. degree in engineering, electromagnetic theory, from Lund University, Lund, Sweden, in 1996 and 2001, respectively. In 2001, he joined the Electromagnetic Theory Group, Department of Electrical and Information Technology, Lund University, where he was appointed Docent in electromagnetic theory in 2005, and is presently a Professor and the Director of Studies of the Department of Electrical and Information Technology. His research interests include electromagnetic properties of materials, composite materials, homogenization, periodic structures, numerical methods, radar cross section, wave propagation in complex and nonlinear media, and the inverse scattering problem.
\end{IEEEbiographynophoto}

\end{document}